\documentstyle[12pt,epsf]{article}

\def\void{}
\def\labelmark{}

{\ifx\void\labelname\def\junk{\end{displaymath}}
\else\def\junk{\end{equation}}\fi\junk\labelmark\def\labelname{}}

{\ifx\void\labelname\def\junk{\end{array}\end{displaymath}}
\else\def\junk{\end{array}\right.\end{equation}}
\fi\junk\labelmark\def\labelname{}\def\junk{}
}

\newcommand{\e}{\mbox{e}}
\newcommand{\mi}{\!-\!}
\newcommand{\equ}{\!=\!}

\begin{document}

\hfill AEI-2001-020

\hfill 19 Apr 2001

\begin{center}
\vspace{80pt}
{ \Large \bf A proper-time cure for the conformal sickness in quantum gravity}

\vspace{60pt}

{\sl A. Dasgupta {\footnote {dasgupta@aei-potsdam.mpg.de}}} and
{\sl R. Loll {\footnote {loll@aei-potsdam.mpg.de}}}

\vspace{36pt}

Max-Planck-Institut f\"{u}r Gravitationsphysik,\\
Am M\"uhlenberg 1, D-14476 Golm, Germany\\

\vspace{48pt}

\end{center}
\vspace{1.5cm}

\begin{center}
{\bf Abstract}
\end{center}
Starting from the space of Lorentzian metrics,
we examine the full gravitational path integral in 3 and 4
space-time dimensions. 
Inspired by recent results obtained in a regularized, dynamically
triangulated formulation of Lorentzian gravity, 
we gauge-fix to proper-time 
coordinates and perform a non-perturbative ``Wick rotation''
on the physical configuration space. 
Under certain assumptions about the behaviour of the 
partition function under renormalization, we find that the
divergence due to the conformal modes of the metric is
cancelled {\it non-perturbatively} by a Faddeev-Popov
determinant contributing to the effective measure. 
We illustrate some of our claims by a 3d perturbative calculation.

\vspace{1cm}


\vspace{12pt}
\noindent


\newpage

\section{Path integrals for quantum gravity}

It is the central aim of path integral formulations of quantum gravity 
to give a physical and mathematical meaning to the formal expression 
\begin{equation}
Z= \int_{\frac{\rm Metrics(M)}{Diff(M)}} 
{\cal D}[g_{\mu\nu}] \e^{i S},\qquad 
S[g_{\mu\nu}]=\frac{1}{16\pi G_N}\int_M d^4x \sqrt{|\det g|}
(R-2\Lambda),
\label{prop}
\end{equation}
for the gravitational propagator, subject to boundary conditions
on the metric fields $g_{\mu\nu}(x)$. The earliest attempts to
construct a Feynman propagator for gravity \cite{mis,leut} 
go back to a time when neither of the present authors had been born 
or, well, barely.
The perturbation series for (\ref{prop}) around flat Minkowski 
space $\eta_{\mu\nu}$ is non-renormalizable and thus cannot 
serve as a {\it fundamental} definition of the theory.
Assuming that a quantum theory of gravity does indeed exist, 
one is therefore forced to consider non-perturbative methods for
constructing $Z$. However, a
non-perturbative evaluation of (\ref{prop}) in the continuum 
meets with a number of well-known problems: 
\begin{itemize}
\item[(i)] explicit field coordinates on the physical configuration space 
$\frac{\rm Metrics(M)}{\rm Diff(M)}$ of diffeo\-morphism-equivalence classes
of metrics $[g_{\mu\nu}]$ (the so-called {\it geometries}) must be
found; 
\item[(ii)] a measure ${\cal D}[g_{\mu\nu}]$ on the ``space of paths''
(the set of all $d$-dimensional space-time geometries
interpolating between an initial and a final spatial geometry)
must be given;
\item[(iii)] since we are dealing with a field theory, a
regularization and renormalization -- respecting the diffeomorphism
symmetry of the gravitational theory -- must be found.
\end{itemize}
\noindent Even if good candidates (i)-(iii) have been identified,
we still expect difficulties with the evaluation of the non-perturbative 
integral since
\begin{itemize}
\item[(iv)] the action is not quadratic in the fundamental metric fields; 
\item[(v)] the integral is unlikely to converge
because of the imaginary factor $i$ in front of the Einstein action.
\end{itemize}

In ordinary quantum field theory on a fixed Minkowskian background, 
problem (v) is
usually solved by rotating to imaginary ``time'', evaluating n-point
functions in the Euclidean sector and invoking the
Osterwalder-Schrader axioms. It is much less obvious how to proceed in
gravity, where the metric field is a dynamical variable. A generic
metric $g_{\mu\nu}$ has no geometrically distinguished notion of time $t$,
and it is therefore unclear how to perform a Wick rotation of the form 
$t\mapsto \tau\equ it$. 

This difficulty has motivated some researchers to change 
the configuration space of the theory, from
Lorentzian space-time metrics $g_{\mu\nu}$ with signature $(-+++)$ to 
Euclidean metrics $g^{\rm eu}_{\mu\nu}$ with signature $(++++)$, 
and simultaneously to replace the complex
amplitudes $\exp iS[g_{\mu\nu}]$ by real Boltzmann weights 
$\exp -S[g^{\rm eu}_{\mu\nu}]$. It is important to realize that
this substitution is ad hoc in the sense of replacing one physical
problem by another one which -- without a non-perturbative
generalization of the Wick rotation -- is a priori unrelated and 
potentially inequivalent.
 
Unfortunately, because of the so-called ``conformal-factor problem'',
such a procedure does {\it still} 
not guarantee the convergence of the regularized path integral. 
This property is visualized most easily by decomposing the (Euclidean or
Lorentzian) metric $g$ into a product of a conformal factor and a 
metric $\bar g$ of constant curvature according to
$g_{\mu\nu} =\e^{2\lambda}\bar g_{\mu\nu}$.
Rewriting the Einstein action as a function of $\lambda$ and $\bar g$,
the kinematic term $\sim (\nabla_0 \lambda)^2$ 
for the conformal field is seen to contribute with the wrong sign,
making the action unbounded from below, and the functional
$\lambda$-integration in the Euclidean
case potentially divergent.

This ``conformal sickness'' has been known 
since the early days of the Euclidean path integral \cite{gibb,hawk}.
Following the suggestion of performing a ``conformal rotation'' 
$\lambda\mapsto i\lambda$ \cite{gibb} for asymptotically Euclidean 
metrics (and $\Lambda\equ 0$), 
the typical cure consists in a suitable integration over {\it complex} 
instead of real metrics $g_{\mu\nu}$. 
A place where Euclidean amplitudes are essential is the no-boundary proposal 
of Hartle and Hawking \cite{haha}. Extensive studies of cosmological 
models with compact slices have been conducted in search of definite
prescriptions of complex integration contours, satisfying certain
criteria of physicality and semi-classicality \cite{halo,jhjh}. 
For simple mini-superspace models such contours can be found, but it
seems very difficult to come up with a prescription for
selecting a contour {\it uniquely} which at the same time 
could claim some generality. 

Other authors, again in a perturbative context, have insisted that
the proper physical starting point for any analysis should be 
{\it Lorentzian} gravity. Either by working ``backwards'' from
a continuum phase-space path integral in terms of
reduced, physical variables \cite{schleich} or by gauge-fixing
the configuration-space path integral and properly including 
the ensuing Faddeev-Popov determinants \cite{mm},
they have argued that the conformal divergence is spurious. 
These arguments highlight the potential importance of the
measure ${\cal D}[g_{\mu\nu}]$ in (\ref{prop}), an issue to which
we will return in due course.

Because of the ill-definedness of the perturbative path integral,
the relevance of these considerations for a full theory of
quantum gravity is not immediately clear.
To our knowledge, the conformal problem of the path integral 
has not been addressed 
in a genuinely non-perturbative setting. This has to do 
with the general lack of regularizations for gravity within which
this issue could be treated in a mathematically meaningful
way. In addition, going beyond the perturbative case, a
Wick rotation on the space of all metrics is needed if one
believes -- as we do -- that the Lorentzian signature and
the causal structure of space-time are of
fundamental physical importance, and should therefore be built
into any quantization from the outset.

Our interest in gravitational path integrals is motivated by the 
recent construction of a non-perturbative regularized path integral for 
gravity based on simplicial Lorentzian geometries \cite{al,ajl1,ajl2}
(see \cite{review} for recent reviews), 
a Lorentzian version of previously investigated so-called
dynamically triangulated models. The model can be defined in
any dimension $d$, possesses a well-defined notion
of Wick rotation and a set of causality constraints reflecting
the properties of the discrete Lorentzian structure. 

This formulation of quantum gravity 
goes some way in addressing the list of problems mentioned
earlier. In the spirit of Regge's old idea of describing ``geometries
without coordinates'' \cite{regge},
it is defined directly on the physical space of geometries.
The (discretized) geometries are
described in terms of the combinatorial data of how a set of 
flat $d$-dimensional simplicial building blocks (whose metric properties 
are encoded in their geodesic edge lengths) is glued together.
This amounts to a definite prescription for (i)-(iii) above. The 
non-perturbative Wick rotation gets rid of the factor of $i$ of problem 
(v), and the Wick-rotated path integral can be shown to converge for a 
suitable choice of bare coupling constants. 

It is remarkable that a regularization for quantum gravity with
such properties should exist and it is of great interest
to understand whether the path integral can be evaluated explicitly,
and simultaneously the diffeomorphism-invariant cut-off be removed
to give rise to a well-defined continuum theory. Instead of
performing Gaussian continuum integrals as in (iv), 
``solving the model'' means the evaluation of 
the discrete combinatorial state sum over distinct gluings.   
This program can be carried out exactly by analytical methods in dimension 
$d\equ 2$ \cite{al}, yielding a well-defined propagator
(\ref{prop}), in agreement with a (formal) continuum calculation in
proper-time gauge \cite{nakayama}. 

These results are reassuring as far as the consistency of 
Lorentzian dynamically triangulated gravity is concerned,
but more serious problems are expected to appear 
in higher dimensions, in the case of the conformal-factor
problem for $d\geq 3$. Although the discrete model always
possesses a phase where $Z$ converges, this may be
attributed to the effective curvature bounds inherent in
the regularization. It does not necessarily exclude 
a dominance of the unphysical conformal mode in the state sum. 
One can indeed identify simplicial geometries (whose
spatial volumes oscillate rapidly in proper time) with a
large and negative Euclidean action. Nevertheless,
it has been established by numerical simulations
that for the 3d model there is a large range
of the gravitational coupling constant where such modes
do not play a role \cite{ajl3d,ajl3dl}. 
This entails a win of ``entropy over energy'',
that is, well-behaved geometries outnumber completely the
potentially dangerous ones associated with conformal
excitations. 

It would be very significant if the same behaviour persisted
in four space-time dimensions, since it would suggest a
resolution of the conformal-factor problem {\it at the
non-perturbative level}, where a quantum theory
of gravity has a chance of existing. The question we will
address in the present work is whether and how such a behaviour
can be understood from a continuum point of view. 
The evidence from Lorentzian dynamical triangulations so far
suggests that a crucial contribution in the cancellation of
the conformal divergence must come from the path integral
{\it measure}. 

To imitate the discretized formulation as closely as possible,
we will use a configuration space 
path integral in terms of metric fields $g_{\mu\nu}$. Our
calculations will be done for $d\equ 3,4$.
In order to gauge-fix, we will work with ``proper-time'' (or
``Gaussian'') coordinates. This is motivated by the presence of
a preferred notion of (discrete) proper time in the lattice model
(although it should be pointed out that in this case there is no 
gauge-fixing -- proper time is simply selected from the combinatorial
data characterizing each geometry).

We do not expect to be able to perform the non-perturbative
functional integrals explicitly (this is problem (iv) from above),
but we will show that under certain plausible assumptions about
the behaviour of the path integral under renormalization the
conformal divergence is cancelled by a compensating term
in the measure, arising as a Faddeev-Popov determinant during the
gauge-fixing. Our treatment will concentrate on the conformal
factor-dependence and will remain formal in the sense that
we will not spell out the details of the
regularization and renormalization. However, the results from the
simplicial formulation make us confident that suitable
diffeomorphism-invariant regularization schemes do indeed exist. 

The cancellation mechanism we uncover is a non-perturbative version
of the one found by Mazur and Mottola \cite{mm}, and requires that
$C < -\frac{2}{d}$ for the constant $C$ appearing in the DeWitt
measure, exactly the range where the DeWitt metric is
indefinite. It leads us to conjecture that the ``natural'' measure
given by the dynamical triangulations approach corresponds to a
value of $C < -\frac{2}{d}$. This is quite plausible, given that
the only distinguished value of $C$ (inherent in the action and
appearing explicitly in a canonical treatment
of three- and four-dimensional gravity) is $C\equ -2$, which lies
in the required range. 

The contents of the remainder of our paper is as follows:
in the next section we will separate out the gauge components of the
metric tensor and discuss some
properties of the proper-time gauge. We also introduce our
conventions for various scalar products.
In Sec.3, we explicitly isolate the negative-definite
part of the action responsible for the conformal divergence.
The Faddeev-Popov determinants associated with the variable
changes on the space of metrics are computed in Sec.4. We then
show that under certain assumptions on the renormalization
behaviour of the state sum (borrowed from 2d Liouville gravity), a 
piece of these determinants exactly cancels the leading
conformal divergence in the action. Sec.5 contains a
{\it perturbative} evaluation of the complete proper-time path integral 
around a fixed constant-curvature torus
metric in 3d, to illustrate the cancellation mechanism at work in
a complete and explicit calculation.
In the final Sec.6, we summarize our findings.

\section{Implementing the proper-time gauge}

Our first task will be to split the metric degrees of freedom
into physical and gauge components, and to divide the gravity
partition function by the (infinite) volume of the diffeomorphism
group. We will work with $d$-dimensional space-times $M$, $d\equ 3,4$, 
with topology ${}^{(d)}M=[0,1]\times {}^{(d\mi 1)}\Sigma$,
where $\Sigma$ denotes a compact spatial manifold. 

In an attempt to follow as closely as possible the discrete
construction of \cite{ajl1,ajl2}, we will represent the physical
configuration space of geometries on $M$ (i.e. the quotient space of
space-time metrics ${\cal M}={\rm Metrics}(M)$ and space-time 
diffeomorphisms Diff$(M)$) by
the space of metrics in ``proper-time gauge''\footnote{Note that
this gauge was used by Leutwyler in one of the first path-integral 
treatments \cite{leut}. In the context of the {\it canonical} path 
integral, a similar proper-time gauge was employed in \cite{teit}.}, 
which are of
the block-diagonal form 
\begin{equation}
g^{\rm pt}_{\mu\nu}=
\left( \matrix{\epsilon&\vec 0\cr \vec 0 & g_{ij}\cr } \right),\;\;
\mu,\nu=0,\dots, d\mi 1,\;\; i,j=1,\dots, d\mi 1.
\label{block}
\end{equation}
Our ``Wick rotation'' consists in substituting 
$\epsilon= -1$ in the Lorentzian case by $\epsilon =+1$ in the
Euclidean case (where we define $\sqrt{-1} :=+i$). For the case that
the spatial (d--1)-dimen\-sio\-nal metric $g_{ij}$ is time-independent 
-- as for instance
in the case of the flat Minkowski metric -- this prescription
is equivalent to an analytical continuation in proper time $t$. 
It is not straightforward to define an exact analogue of
the discrete Wick rotation of \cite{ajl1,ajl2}, which is given
as an operation on discrete {\it geometries}, without the need to
introduce any coordinates. In that case, one can nevertheless 
choose a coordinate
system on each individual flat simplicial building blocks in which
the metric tensor takes the form (\ref{block}), and
the discrete Wick rotation (up to a constant rescaling of proper 
time) corresponds to a sign flip of the (00)-component. Another
property our Wick rotation shares with the discrete case is the
fact that it maps real Lorentzian metrics to real Euclidean metrics.
Note that unlike its discrete counterpart, our prescription 
$\epsilon\mapsto -\epsilon$ does not in general map
solutions of Lorentzian gravity to Euclidean solutions. (For the
dynamically triangulated Lorentzian models this is ensured in the sense that
the two actions are mapped into each other.) However, this is no
obstacle to our non-perturbative construction, where 
$\epsilon\mapsto -\epsilon$ simply gives us a 1-to-1 map
from Lorentzian to Euclidean geometries. All computations are then
performed in the Euclidean sector where -- up to regularization --
they are well-defined. We will not address the question of what is
the most suitable way of ``rotating back the result'', since
this will ultimately be dictated by the physical interpretation of the
final, non-perturbative partition function (obvious candidates are
an inverse flip $\epsilon \equ 1\mapsto \epsilon\equ -1$ or an analytic 
continuation in proper time).

One may wonder why we have not adopted a prescription of the
form of an analytic continuation in time, $t\mapsto it$.
The problem is that although such prescriptions ``work'' for a
handful of metrics $g_{\mu\nu}$ with special symmetries (flat space,
static solutions etc.), they do not exist in general. 
Firstly, a generic space-time does not have a physically preferred 
time-direction, and the prescription is clearly not invariant under
diffeomorphisms. Secondly, if by some gauge choice one does 
distinguish a preferred system of coordinates (like the Gaussian
coordinates we are using), the substitution $t\mapsto it$ will
in general lead to {\it complex} metrics, defeating the purpose
of making the non-perturbative path integral better defined.

Keeping track of the signature is particularly convenient 
in proper-time gauge and we will work throughout 
with factors of $\epsilon$.
The metric $g_{ij}$ is taken to be
positive definite. Locally on a space-time one can always 
find so-called ``Gaussian normal coordinates'' in which the
metric tensor assumes the form (\ref{block}), 
but in general one
expects obstructions to introducing such coordinates globally.

As with any gauge choice the gauge must be {\it attainable} and
{\it unique}, that is, any point $g_{\mu\nu}\in\cal M$ must lie
on a gauge orbit that cuts the constraint surface $\cal C$ 
(in our case defined by the vanishing of the gauge condition,
$g_{0\mu}-\epsilon \delta_{\mu}^{0}\equ 0$) exactly once.
A necessary condition which is easier to prove
is that any $g_{\mu\nu}$ in the vicinity of $\cal C$
can be uniquely decomposed into a configuration $g_{\mu\nu}^{\rm pt}
\in\cal C$ and an infinitesimal diffeomorphism. This is
demonstrated in appendix 1. 

\begin{figure}
\centerline{ \epsfxsize 3in
\epsfbox{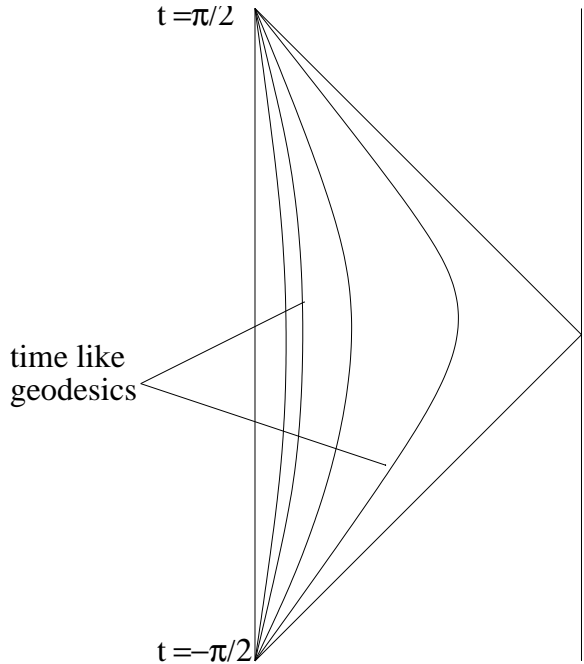}
}
\caption{\sl Penrose diagram of anti-de Sitter space, illustrating the
convergence of time-like geodesics.
}
\label{fig:ades}
\end{figure}

Potential difficulties with the global implementation of the
proper-time gauge have to do with the focussing properties of
time-like geodesics. Anti-de Sitter space in 3 and 4 dimensions is
an example of a solution to the classical Einstein equations 
where proper-time coordinates do not cover the
entire space-time. The 4d metric in these coordinates assumes the form 
\begin{equation}
ds^2 = - dt^2 + \cos^2t \left( d\chi^{2} + \sinh^2\chi (d\theta^2
+\sin^{2}\theta d\phi^{2} ) \right),
\end{equation}
with coordinate singularities at $t=\pm \pi/2$, where the time-like
geodesics orthogonal to the hypersurfaces $t=const$ converge to points,
as is illustrated by the Penrose diagram in Fig.\ \ref{fig:ades}. 
This is therefore an example of a metric --
albeit one with rather bizarre causality properties \cite{hawkellis} --
that cannot be reached from the constrained surface $\cal C$ by
a diffeomorphism. (By contrast, no such problems occur
for the {\it de Sitter} space, say.)

The existence of such configurations in an
infinite-dimensional context is not surprising. 
For example, it is well-known 
from the Riemannian case \cite{fischer} that configurations
$g_{\mu\nu}$ with special symmetries must be excised from $\cal M$
to make quotient spaces of the kind ${\cal M}/{\it Diff}(M)$
well-defined. Similarly,
for the purposes of the non-perturbative path
integral we are only interested in capturing the properties of ``generic''
metrics, and not of ``sets of measure zero''.
In our work we assume that the diffeomorphism orbits 
$f^{*}g^{\rm pt}_{\mu\nu}$ through metrics of the special form (\ref{block})
are in a suitable, function-theoretic sense dense in the space $\cal 
M$ of all metrics.\footnote{As far as we can see, this assumption 
is not in contradiction with the well-known
difficulty of using proper-time coordinates in an {\it initial-value}
formulation of the classical Einstein equations: in this case, given 
initial data for the spatial metric $g_{ij}$ and
the extrinsic curvature $K_{ij}$, caustics in the proper-time
coordinate system will generically develop at some finite 
(positive or negative) time if
${\rm Tr}\ K\not\equiv 0$ initially (see, for example, \cite{land}).}
Since we do not have a precise definition of a suitable 
quantum analogue
of the space ${\cal M}/{\it Diff}(M)$ beyond formal continuum
calculations, such an assumption can ultimately only be justified 
by the results of a properly regularized formulation of the theory.

We now must implement our gauge choice to isolate
the physical degrees of freedom. This requires a change of
coordinates $g_{\mu\nu}\mapsto (g^{\rm pt}_{\mu\nu},f)$
on $\cal M$, where $f$ labels space-time diffeomorphisms that map
any boundaries of $\cal M$ into themselves. 
(We do not specify any detailed boundary conditions
because our main argument will not depend on them.)
Such a coordinate transformation
must be accompanied by a Jacobian \cite{leut,fp}, whose explicit 
functional form
depends on the gauge condition imposed on the metric. We will 
determine this Jacobian in proper-time gauge using the methods of
Mottola et al \cite{mm,bbm} (see also \cite{mottola} for a
pedagogical introduction). We decompose an arbitrary tangent
vector $\delta g_{\mu\nu}|_{g}\equiv h_{\mu\nu}|_{g}$ 
in a point $g\in\cal M$ according to
\begin{equation}
h_{\mu\nu}=
h_{\mu\nu}^{\rm pt}+\nabla_{\mu}\xi_{\nu}+\nabla_{\nu}\xi_{\mu}=:
h_{\mu\nu}^{\rm pt}+(L\xi)_{\mu\nu},
\label{decomp}
\end{equation}
where $h_{\mu\nu}^{\rm pt}$ is the gauge-invariant piece of 
$h_{\mu\nu}$ defined by
\begin{equation}
(F\circ h^{\rm pt})_{\mu}=F^{\nu}h_{\mu\nu}^{\rm pt}\equiv
\delta_{0}^{\nu}h_{\mu\nu}^{\rm pt}=
h_{\mu 0}^{\rm pt}=0,
\label{gauge}
\end{equation}
and the vector field $\xi$ generates an infinitesimal diffeomorphism
of $M$. Note that we are not separating out the
trace-free part of the metric at this stage.
A natural scalar product for tangent vectors to $\cal M$ is given by 
\begin{equation}
\langle h,h' \rangle^{\rm T}|_{g} =\int_{M}d^{d}x\ \sqrt{|\det g |}\
h_{\mu\nu} G^{\mu\nu\rho\sigma}_{(C)}
h'_{\rho\sigma},
\label{scalart}
\end{equation}
where the ``T'' stands for ``tensor'' and we will from now on suppress
the dependence on the base point $g\in \cal M$. The DeWitt metric is
\begin{equation}
G^{\mu\nu\rho\sigma}_{(C)}=\frac{1}{2} ( g^{\mu\rho}g^{\nu\sigma}+
g^{\mu\sigma}g^{\nu\rho}+C g^{\mu\nu}g^{\rho\sigma}),
\label{dewitt}
\end{equation}
for an arbitrary real constant $C$. Similarly, the scalar product for
vector fields on $M$ is
\begin{equation}
\langle \xi,\xi'\rangle^{\rm V}=\int_{M}d^{d}x\ \sqrt{|\det g|}\
\xi_{\mu} g^{\mu\nu} \xi'_{\nu},
\label{scalarv}
\end{equation}
and for scalars
\begin{equation}
\langle \omega,\omega'\rangle^{\rm S}=\int_{M}d^{d}x\ \sqrt{|\det g|}\
\omega \omega'.
\label{scalars}
\end{equation}
Note in passing that Lorentz-invariance is not an issue in defining
expressions like (\ref{scalart},\ref{scalarv},\ref{scalars}), since the
metric manifold $(M,g_{\mu\nu})$ does not carry a global action
of the Lorentz group unless $g$ is the flat Minkowski metric.
(It {\it would} be a requirement if we were considering
a perturbative formulation around flat space.)
On the other hand, diffeomorphism invariance of the 
entire path integral will be maintained throughout our construction.  

The (base-point dependent) Jacobian $J_\epsilon[g_{\mu\nu}]$ is defined by
\begin{equation}
[{\cal D} h_{\mu\nu}]_\epsilon=J_\epsilon\  
[{\cal D} h_{\mu\nu}^{\rm pt}]_\epsilon [{\cal 
D}\xi_{\mu}]_\epsilon,
\label{jacob}
\end{equation}
and can be computed by assuming Gaussian normalization conditions of
the form
\begin{equation}
\int [{\cal D} h_{\mu\nu}]_\epsilon \exp \left[ -\frac{\sqrt{\epsilon}}{2}
\langle h,h \rangle_{\epsilon}^{\rm T} \right] =1,
\;\;\; \int [{\cal D} \xi_{\mu}]_{\epsilon} \exp 
\left[ -\frac{\sqrt{\epsilon}}{2}
\langle \xi,\xi \rangle_{\epsilon}^{\rm V} \right] =1
\label{norm}
\end{equation}
and similarly for $h_{\mu\nu}^{\rm pt}$. 
The diffeomorphism-invariance of the measure $[{\cal D} h_{\mu\nu}]$
has been shown in \cite{bbm}. 
We have introduced the subscript
$\epsilon$ for the measures and scalar products 
to indicate their dependence on
the signature. Analogously, functional determinants of suitable
operators $\cal O$ are computed according to
\begin{equation}
\int [{\cal D} h_{\mu\nu}]_\epsilon \exp \left[ -\frac{\sqrt{\epsilon}}{2}
\langle h,{\cal O} h \rangle_{\epsilon}^{\rm T} \right] =
\frac{1}{\sqrt{\det {\cal O}}}.
\label{op}
\end{equation}
The way the $\epsilon$-dependence is to be interpreted in the
functional integrals above is as follows. All computations are to
be performed for the Euclidean value $\epsilon\equ 1$ and then continued. 
This continuation can be non-trivial in a relation like
(\ref{op}) only if the original operator $\cal O$ had an explicit
$\epsilon$-dependence.

The measure $[{\cal D}g_{\mu\nu}]$ for the full path integral
(\ref{prop}) must be diffeomorphism-in\-va\-riant and is usually
assumed to be closely related to the tangent space measure
$[{\cal D}h_{\mu\nu}]$. 
Since both the measure and the Einstein action are invariant under 
diffeomorphisms,
we can factor out the volume of the d-dimensional diffeomorphism group
and are left with the path integral (c.f. \cite{mm,bbm})
\begin{equation}
Z^{(\epsilon)}=\int [{\cal D} g^{\rm pt}_{\mu\nu}]_{\epsilon}\ 
J_{\epsilon}[g^{\rm pt}]\ 
\e^{i \sqrt{-\epsilon }S_{\epsilon}[g^{\rm pt}]}.
\label{partlor}
\end{equation}
The gravitational action in terms of the gauge-fixed metrics 
appearing in (\ref{partlor}) is given by
\begin{equation}
S_{\epsilon}[g]=
-\frac{\epsilon}{16\pi G}\int d^dx \sqrt{\det g_{ij}}
\left({}^{(d-1)}R -2\Lambda - \frac{\epsilon}{4}
G^{ijkl}_{(-2)} (\partial_0 g_{ij})(\partial_0 g_{kl})
\right),
\label{actfix}
\end{equation}
where ${}^{(d-1)}R$ denotes the scalar curvature of the spatial
metric $g_{ij}^{\rm pt}$, and where we
have now dropped the explicit superscript indicating proper-time gauge.
The Jacobian has the form
\begin{equation}
J= 
\sqrt{{\det}_{\rm V}(F\circ F^{\dag})^{-1}(F\circ L)^{\dag}(F\circ L)},
\label{deter}
\end{equation}
where $L$ -- defined in (\ref{decomp}) -- 
maps vectors to symmetric tensors, and 
$F$ -- the gauge condition according to (\ref{gauge}) --
maps symmetric tensors into vectors. We tacitly assume that
zero eigenvectors have been removed in the computation of determinants
like (\ref{deter}).
Adjoints and determinants are defined with respect to the
scalar product on $d$-vectors induced by the DeWitt metric 
(\ref{dewitt}) at $g=g^{\rm pt}$, namely,
\begin{equation}
\langle \vec\eta,\vec\epsilon\rangle =\int d^{d}x 
\sqrt{\det g^{\rm pt}_{ij}}
\ \eta_{\mu}^{*}\ (g^{\rm pt})^{\mu\nu}\ \epsilon_{\nu}.
\label{vecprod}
\end{equation}

\section{Isolating the conformal divergence}

As we have already described in the introduction, the Euclidean gravity 
path integral in $d\geq 3$ suffers from a ``conformal sickness'' arising 
because of the unboundedness of the action from below
\cite{gibb,mm}.
It is straightforward to see 
that the same is true for the action (\ref{actfix}) in proper-time
gauge. To isolate the relevant kinetic terms, we 
decompose the time-derivatives according to
\begin{eqnarray}
\partial_0 g_{ij} &=& (\partial_{0}g_{ij})^{\Vert} +
(\partial_{0}g_{ij})^{\perp} \nonumber\\
&:=&
(1-\tilde G)_{ij}{}^{kl} (\partial_0 g_{kl}) +
\tilde G_{ij}{}^{kl} (\partial_0 g_{kl}),
\label{timeder}
\end{eqnarray}
into a trace part and a trace-free part, where the projector $\tilde G$
onto the trace-free subspace is given by
\begin{equation}
\tilde G_{ij}{}^{kl} =\frac{1}{2} (\delta_i{}^l\delta_j{}^k +
\delta_i{}^k\delta_j{}^l -\frac{2}{d-1}\ g_{ij}\ g^{kl}).
\label{project}
\end{equation}
The kinetic term in (\ref{actfix}) can be rewritten as
\begin{eqnarray}
G^{ijkl}_{(-2)} (\partial_0 g_{ij})(\partial_0 g_{kl})&=&
-\frac{d-2}{d-1}\ 
(\partial_{0}g_{ij})^{\Vert} g^{ij}g^{kl}(\partial_{0}g_{kl})^{\Vert}
+(\partial_{0}g_{ij})^{\perp} g^{ik}g^{jl}(\partial_{0}g_{kl})^{\perp}
\nonumber\\
&=& -\frac{d-2}{d-1}\ \frac{(\partial_0\det g )^2}{(\det g)^2} +
(\partial_0 g_{ij})^{\perp} g^{ik}g^{jl}(\partial_0 g_{kl})^{\perp}.
\label{traces}
\end{eqnarray}
The first term on the right-hand side is the negative definite 
trace-part. This is precisely the kinetic term 
of the conformal mode  
one isolates in perturbative expansions around Ricci-flat 
metrics, and which leads to the conformal divergence. 
The second term in ({\ref{traces}), coming from the trace-free
directions, is positive definite (c.f. \cite{dewitt}).
In order to make the $\lambda$-dependence more explicit,
we decompose the metric according to $g_{ij}=\e^{2\lambda} \bar g_{ij}$,
where $\bar g_{ij}$ is a constant-curvature metric. This is always
possible for the Riemannian metrics we are considering \cite{schoen},
but note that in our case $g_{ij}$ has an additional time-dependence.

We will deal with the Jacobian accompanying
the coordinate change $g_{ij}\mapsto (\lambda,\bar g_{ij})$ 
in the next section. The complete action (\ref{actfix}) becomes
\begin{eqnarray}
&&S_{\epsilon }=
-\frac{\epsilon }{16\pi G}\int d^dx \sqrt{ \det\bar g_{ij} }\ 
\Biggl(
\e^{(d-3)\lambda } [\bar {R} + (d-2)(d-3)(\bar\nabla_{i}\lambda )
(\bar\nabla^{i}\lambda) ]  +
\nonumber\\
&& \hspace{.2cm}\e^{(d-1)\lambda} \Bigl( -2\Lambda +\epsilon (d-1)(d-2)
[ \partial_{0}(\lambda +\frac{\log\det\bar g}{2 (d-1)})]^{2}-
\frac{\epsilon}{4} (\partial_{0}\bar g_{ij})^{\perp} \bar g^{ik}
\bar g^{jl} (\partial_{0}\bar g_{kl})^{\perp}) \Bigr)  \Biggr) ,\nonumber\\
\label{kinet}
\end{eqnarray}
and is unbounded below (for either signature) because of the
``wrong'' sign for the kinetic term in the shifted scaling parameter
\begin{equation}
\tilde\lambda= \lambda + \frac{1}{2(d-1)}\log {\det\bar g}.
\label{lamtilde}
\end{equation}
This presents a potential problem for the Euclidean
approach where the 
exponentiated action contains a term
$\sim \e^{\gamma^2 \int (\partial_{0} \tilde\lambda )^{2}}$ which can become
arbitrarily large for strongly oscillating conformal factors.
For later convenience, we rewrite the action in terms of
the shifted variable $\tilde\lambda$ as
\begin{eqnarray}
&&S_{\epsilon }=
-\frac{\epsilon }{16\pi G}\int d^dx\ \Biggl( (\det\bar g_{ij} )^{\frac{1}{d-1}} 
\e^{(d-3)\tilde\lambda } [\bar {R} + (d-2)(d-3)(\bar\nabla_{i}\tilde\lambda )
(\bar\nabla^{i}\tilde\lambda) ]  +
\nonumber\\
&& \hspace{.3cm}\e^{(d-1)\tilde\lambda} \Bigl( -2\Lambda +\epsilon (d-1)(d-2)
[ \partial_{0}\tilde\lambda]^{2}-
\frac{\epsilon}{4} (\partial_{0}\bar g_{ij}) (\bar g^{ik}
\bar g^{jl}-\frac{1}{(d-1)} \bar g^{ij} \bar g^{kl})
 (\partial_{0}\bar g_{kl}) ) \Bigr) \Biggr),\nonumber\\
\label{actnew}
\end{eqnarray}
where we have now written out the positive-definite
kinetic term explicitly.
Note also that $S_{\epsilon}$ is non-polynomial in both $\tilde\lambda$
and $\bar g_{ij}$. In three dimensions, the expression (\ref{actnew})
simplifies further to
\begin{eqnarray}
&&S_{\epsilon }^{d=3}=
-\frac{\epsilon }{16\pi G}\int dt \Bigl( 4\pi\chi +
\int d^2x\ \e^{2\tilde\lambda} \bigl( -2\Lambda +2 \epsilon 
[ \partial_{0}\tilde\lambda]^{2}\nonumber\\ 
& &\hspace{2cm} -\frac{\epsilon}{4} (\partial_{0}\bar g_{ij}) (\bar g^{ik}
\bar g^{jl}-\frac{1}{2} \bar g^{ij} \bar g^{kl})
 (\partial_{0}\bar g_{kl}) \bigr)  \Bigr),
\label{act3}
\end{eqnarray}
with $\chi$ denoting the Euler characteristic of the two-dimensional
spatial manifold.

From the point of view of the canonical formulation of gravity
the presence of the conformal divergence is rather puzzling. 
In that case it is clear that the conformal factor is not
a propagating degree of freedom, since it is canonically
conjugate to a gauge variable and becomes fixed by imposing the Hamiltonian 
constraint. In metric path integrals of the kind we are considering, this 
property is not at all obvious. The
natural place to look for a cancellation of the divergence is the 
path-integral measure, which is a central ingredient in
any non-perturbative formulation. 
As we have mentioned earlier, this scenario seems to be realized
in the non-perturbative approach based on piece-wise linear Lorentzian 
geometries \cite{al,ajl1,ajl2}, which is 
one of the few well-defined regularized path integrals available
that do not rely on any fixed background geometry. 
Numerical investigations of the corresponding continuum theory
in $d\equ 3$ have shown that for sufficiently
large bare Newton constant there is a phase whose ground state has
a stable and extended geometry, without the large 
fluctuations indicative of conformal excitations \cite{ajl3d}. 

This clearly non-perturbative effect has motivated us to re-examine
the continuum gravitational path integral, to understand how such
a cancellation may occur when the measure is properly taken into
account. Having identified the explicit form of the conformal divergence
in proper-time gauge, we will now look for potentially compensating
terms in the measure, more precisely, relevant contributions in the form of
Faddeev-Popov determinants.

\section{Computing the measure and cancelling the divergence}

Next we determine the measure contributions arising as a result of
the coordinate transformations of the previous section,
$g_{\mu\nu}\mapsto (g^{\rm pt}_{\mu\nu},f)$ and
$g^{\rm pt}_{\mu\nu}\equiv g_{ij}\mapsto (\bar g_{ij},\tilde\lambda)$.

The two functional determinants appearing in the Jacobian $J$
in (\ref{deter}) are vector determinants. The operators
in their arguments are formally self-adjoint, because they
are of the form of products of operators with their adjoints. 
Computing the explicit operators, one finds 
\begin{equation}
(F^\dag\xi)_{\mu\nu}=\frac{1}{2} (\xi_\mu g_{\nu 0}+ \xi_\nu g_{\mu 0})
-\frac{C}{d C+2}\ \xi_0 g_{\mu\nu}
\label{fadjoint}
\end{equation}
for the adjoint of $F$, leading to the diagonal operator
\begin{equation}
(F\circ F^{\dag})_{\mu}{}^{\nu}=\frac{1}{2}\ 
\epsilon \delta_{\mu}{}^{\nu} +\frac{C (d-2)+2}{2 (d C+2)}\ g_{\mu 
0}\delta_{0}{}^{\nu}
=\frac{\epsilon}{2}
\left( \matrix{\frac{2 (d-1) C+4}{d C+2} &\cr &1_{d-1}\cr}
\right),
\label{fdet}
\end{equation}
where $1_{d-1}$ denotes the ($d\mi 1$)-dimensional unit matrix.
We will be needing the determinant of the inverse of this operator 
which for later convenience we factorize into a scalar and a
spatial ($d\mi 1$)-dimensional vector determinant according to
\begin{equation}
{\det}_{\rm V}(F\circ F^{\dag})^{-1}={\det}_{\rm S} \left( 
\frac{2+dC}{2+(d-1)C} \right) \ 
{\det}_{{\rm V}^{d-1}} \left( 2\epsilon \right) .
\end{equation}

The remaining terms in the Jacobian $J$ depend on the operator
\begin{equation}
(F\circ L)_\mu{}^\nu=\delta_0{}^\nu\nabla_\mu +\delta_\mu{}^\nu
\nabla_0
\label{fl}
\end{equation}
together with its adjoint,
\begin{equation}
(F\circ L)^\dagger_\mu{}^\nu=-g_{0\mu}\nabla^\nu -\Gamma_{0\mu}^\nu
-\delta_\mu{}^\nu (\nabla_0+\Gamma_{0\lambda}^\lambda).
\label{fladj}
\end{equation}
Substituting in the expression 
$\nabla_0^\dagger =-(\nabla_0 +\Gamma_{0\mu}^\mu)$
for the adjoint of $\nabla_0$, we finally obtain
\begin{eqnarray}
[(F\circ L)^{\dag}(F\circ L)]_{\mu}{}^{\nu} &=&
-g_{0\mu}\delta_{0}{}^{\nu}\nabla^\lambda\nabla_\lambda
-\delta_{0}{}^{\nu} \Gamma_{0\mu}^\lambda \nabla_\lambda+
\delta_{0}{}^{\nu} \nabla_0^\dagger \nabla_\mu 
\nonumber\\ 
&&-g_{0\mu}\nabla^{\nu}\nabla_{0} - 
\Gamma_{0\mu}^\nu \nabla_0 
+ \delta_{\mu}{}^{\nu}
\nabla_0^\dagger\nabla_0.
\label{fldet}
\end{eqnarray}

Note that the determinant of this operator can be written as a
product of two determinants of operators which are separately 
self-adjoint, namely,
\begin{eqnarray}
{\det}_{\rm V} (F\circ L)^{\dag}(F\circ L) &=&
{\det}_{\rm V} (\nabla_{0}^{\dagger}\nabla_0)\
{\det}_{\rm V} ((\nabla_0^{\dagger})^{-1}(F\circ L)^{\dag}(F\circ L)
\nabla_0^{-1})
\nonumber\\
&=& {\det}_{\rm V} (\nabla_{0}^{\dagger}\nabla_0)\
{\det}_{\rm V} (F\circ L\circ \nabla_0^{-1})^{\dag}
(F\circ L\circ\nabla_0^{-1})\nonumber\\
&=:& {\det}_{\rm V} (\nabla_{0}^{\dagger}\nabla_0)\ 
{\det}_{\rm V}(K^{\dag}\circ K).
\label{morefldet}
\end{eqnarray}
We have separated out the time derivatives
since we are particularly interested in identifying terms that can
cancel the divergence associated with the conformal kinetic terms in
(\ref{kinet}), (\ref{actnew}).
The Faddeev-Popov operator (\ref{fldet}) contains
terms of the same kind, coming from eigenfunctions $\rho_\nu (x)$
that are rapidly oscillating in time. In the region where
$|\nabla_i\rho_\nu|\ll |\nabla_0\rho_\nu|$, this behaviour 
is captured by the factorized operator $(\nabla_{0}^{\dagger}\nabla_0)$.
The factorization (\ref{morefldet}) 
will be used in the cancellation argument below.

We proceed similarly for the second Jacobian $\tilde J$, which comes
from isolating the divergent mode $\tilde\lambda$ in the action
(c.f. the discussion in Sec.\ 3). For
the purposes of this section, it is not necessary to specify 
explicitly which variables $(g_{ij})^{\perp}$ are used on the
remainder of the configuration space. Using the projectors $\tilde G$
and $1-\tilde G$ as in (\ref{timeder}), (\ref{project}), we
decompose the tangent vectors as
\begin{equation}
\delta g_{ij} = (\delta g_{ij})^{\Vert} +
(\delta g_{ij})^{\perp},
\label{tangent}
\end{equation}
where the trace part is given by
\begin{equation}
(\delta g_{ij})^{\Vert}=(1-\tilde G)_{ij}{}^{kl}\delta g_{kl}=
\frac{1}{d-1} g_{ij}\ \delta\log\det g \equiv 2 g_{ij}\ \delta\tilde\lambda
\label{trace}
\end{equation}
and $\tilde\lambda$ has already appeared in (\ref{lamtilde}).
The Jacobian $\tilde J$ is now defined through
\begin{eqnarray}
1&=&\int [{\cal D}\delta g^{\Vert}]_{\epsilon} \int [{\cal D}\delta
g^{\perp}]_{\epsilon}\ 
\e^{-\frac{\sqrt{\epsilon}}{2} \langle \delta 
g, \delta g\rangle_{C} } \nonumber\\
&=&\tilde J \int [{\cal D}\delta \tilde\lambda]_{\epsilon} 
\int [{\cal D}\delta
g^{\perp}]_{\epsilon}\ 
\e^{-\frac{\sqrt{\epsilon}}{2} 
(\langle \delta g^{\Vert}(\delta\tilde\lambda), 
\delta g^{\Vert}(\delta\tilde\lambda)\rangle_{C}+
 \langle \delta g^{\perp}, \delta g^{\perp}\rangle_{C}) },
\label{jac2}
\end{eqnarray}
where the scalar products are taken with respect to the DeWitt metric
(\ref{dewitt}) restricted to the spatial components,
\begin{equation}
\langle \delta g, \delta g\rangle_{C}=\int d^{d}x\ \sqrt{\det g_{ij}}
\ \delta g_{ij}\ G_{(C)}^{ijkl}\ \delta g_{kl}.
\label{restrict}
\end{equation}
As in Sec.\ 2 above, we assume separate Gaussian normalizations for the 
two functional integrals, leading to
\begin{equation}
\tilde J^{-1} =\int [{\cal D}\delta \tilde\lambda]_{\epsilon}\ 
\e^{-\sqrt{\epsilon} (d-1)(2+(d-1)C) \int d^{d}x\
\sqrt{\det g} (\delta\tilde\lambda)^{2}},
\end{equation}
and therefore
\begin{equation}
\tilde J =\sqrt{
{\det}_{\rm S}\ 2 (d-1)(2+(d-1)C)  }.
\end{equation}
We now collect all determinants to obtain
\begin{eqnarray}
J\cdot\tilde J&=&
\sqrt{ {\det}_{\rm S} ((2+dC)(d-1)2\epsilon\partial_{0}^{\dagger}
\partial_0)\;
{\det}_{{\rm V}^{d-1}} ( 2\epsilon 
\nabla_{0}^{\dagger}\nabla_0) \;
{\det}_{\rm V}(K^{\dag}\circ K) },
\label{alldets}
\end{eqnarray}
where we now have decomposed also the vector determinant of the time 
derivatives into a scalar and a spatial vector piece. 
This combined Jacobian appears in our final form of the
non-perturbative continuum path integral in proper-time gauge,
\begin{equation}
Z_\epsilon=\int [{\cal D}g_{ij}^\perp ]_\epsilon
\int [{\cal D}\tilde\lambda ]_\epsilon\ J_\epsilon \tilde
J_\epsilon\ \e^{-\sqrt{\epsilon} S_\epsilon (g_{ij}^\perp,\tilde
\lambda)}.
\label{jacact}
\end{equation}

Unfortunately, but not unexpectedly, there is no immediate way
in either $d\equ 3$ or $d\equ 4$ of evaluating these integrals
since they are not of Gaussian form. 
For the time being we are content with less, namely, understanding
the role played by the conformal factor. For this purpose, let us
concentrate on the leading divergence in $\tilde\lambda$ in the action
(\ref{actnew}),
\begin{eqnarray}
S_D&=& -k\int d^dx\  \e^{(d-1)\tilde\lambda} (\partial_0 \tilde\lambda)^2= 
k\int d^dx\ \e^{(d-1)\tilde \lambda} \tilde\lambda
(\partial_0+\Gamma_{0\mu}^\mu)\partial_0\tilde\lambda\nonumber\\
& =&
-k \int d^dx\ \e^{(d-1)\tilde \lambda} \tilde\lambda \partial_0^\dagger
\partial_0 \tilde\lambda =
-k \int d^dx\ \sqrt{\bar g}\e^{(d-1) \lambda} \tilde\lambda \partial_0^\dagger
\partial_0 \tilde\lambda,
\end{eqnarray}
with the positive constant $k=(d-1)(d-2)/(16\pi G)$, 
neglecting all other terms (including boundary contributions) 
in the action. 

What still stands in the way of our doing the $\tilde\lambda$-integration
in (\ref{jacact}) 
is the $\tilde\lambda$-dependence of the measure
in the action and of the various Jacobians.
Following the example of Distler and Kawai in two-dimensional
Liouville gravity \cite{kawai}, 
we make the unrigorous, but well-motivated
assumption that all measures with respect to the metric 
$g_{ij}=\e^{2\lambda}\bar g_{ij}$ can by suitable field redefinitions
be turned into measures with respect to the constant-curvature
metric $\bar g_{ij}$, where the Jacobian accompanying this
variable change is assumed to
be of the same functional form as the (exponentiated) original 
action\footnote{In principle other (higher-curvature) terms may 
be generated in the process, but they will not affect
our cancellation argument for the conformal divergence.}. 
Applying this philosophy to the truncated path integral, we can 
pull all functional determinants out of the $\tilde\lambda$-integral
(since they are now defined with respect to the metric
$\bar g_{ij}$), resulting in
\begin{eqnarray}
&&Z_\epsilon=\int [{\cal D}g_{ij}^\perp ]_\epsilon
\sqrt{ {\det}_{\rm S} ((2+dC)(d-1)2\epsilon\bar\partial_{0}^{\dagger}
\bar\partial_0)\;
{\det}_{{\rm V}^{d-1}} ( 2\epsilon 
\bar\nabla_{0}^{\dagger}\bar\nabla_0) \;
{\det}_{\rm V}(\bar K^{\dag}\circ \bar K) }\cdot\nonumber\\
&&\hspace{1.5cm}
\int [{\cal D}\tilde\lambda ]_\epsilon\ 
\e^{-\sqrt{\epsilon}\ k^{ren} \int d^dx \sqrt{\bar g}\ 
\tilde\lambda \bar\partial_0^\dagger
\bar\partial_0 \tilde\lambda +\ldots},
\label{trunc}
\end{eqnarray}
where we have absorbed the effect of the new Jacobian in a
renormalization of the gravitational coupling constant contained in
$k$ (and we are assuming that $k^{ren}$ is still 
positive).\footnote{There is a somewhat related path-integral 
treatment by Mazur, based on a conformal 
decomposition of Riemannian space-time metrics \cite{mazur}. 
However, he concentrates on boundary rather than bulk terms in the
effective action.}

Since $\tilde\lambda$ takes values on the entire real line, we can
set ${\cal D}\tilde\lambda\equ {\cal D}\delta\tilde\lambda$ and
perform the $\tilde\lambda$-integral to formally obtain  
$1/\sqrt{{\det}_{\rm S} (-\bar\partial_0^\dagger\bar\partial_0) }$. As can
be seen, this term {\it is} cancelled by the scalar
determinant in (\ref{trunc}) {\it provided} that its prefactor
is negative, that is, if $C <-\frac{2}{d}$. 
Obviously the determinants involved here 
are badly divergent and must in principle be regularized.
However, since the two terms have the same functional form we
expect the cancellation to go through independent of the
regularization chosen. (What we have in mind as a typical 
non-perturbative regularization of the partition function 
(\ref{trunc}) is a common
frequency cutoff $\omega_{n_{0}}$ for the {\it entire} expression
$Z_\epsilon$ (not just for the $\tilde\lambda$-integration).
The regularized determinants then take the form 
${\det}_{\rm S}^{(n_{0})} \hat {\cal O}=\prod_{n\leq n_{0}}e_{n}$,
where $e_{n}$ is the eigenvalue of the n'th eigenfunction of the 
operator $\hat{\cal O}$, and where we are suppressing degeneracies
of the eigenfunctions associated with the spatial directions.)

Thus we conclude that under the plausible assumption that the
conformal part of the volume element $\sqrt{\det g}$ can be
absorbed into a redefinition of the coupling constants and  
provided that the DeWitt metric
is chosen with $C <-\frac{2}{d}$, the conformal divergence of
the Euclidean wick-rotated gravitational path integral is cancelled
{\it non-perturbatively} by a corresponding term in the measure,
coming from a Faddeev-Popov determinant.

The condition $C <-\frac{2}{d}$ was found previously in 
the perturbative treatment of Mazur and Mottola \cite{mm},
with a similar cancellation mechanism.
We differ from their and other authors' treatments
by obtaining the configuration space of ``Euclidean gravity'' 
through a non-perturbative
Wick rotation of the gauge-fixed Lorentzian path integral $Z^{(-1)}$,
expression (\ref{partlor}).
Wick rotation in proper-time gauge corresponds to a 
straightforward substitution $\epsilon\mapsto -\epsilon$ in our
formulas.
In other gauges, there is no immediate relation between
the Euclidean and the Lorentzian sectors {\it beyond the perturbative
regime around a fixed (typically flat) background metric}. 
In these cases, even if any non-perturbative results could be obtained for 
Euclidean signature,
their implications for the physical, Lorentzian sector would be unclear.

By contrast, we have shown under what conditions a cancellation of
the conformal divergence may take place in the full, non-perturbative
theory of Lorentzian space-times. This lends further support to the
finding of the quantum gravity model obtained from Lorentzian 
dynamical triangulations, whose effective measure
(including entropy contributions) apparently leads to a non-perturbative
cancellation of the ``conformal sickness'' of the action. 
The restriction $C<-\frac{2}{d}$ found in the continuum is not unnatural 
in the sense
that this parameter region contains the only dynamically distinguished
value $C\equ -2$ of the constant $C$ (found after 
Legendre-transforming the gravitational action in $d\equ 3$ and 4).

Even with our assumption of the absorption of the scaling factors
$\e^{2\lambda}$, it is unlikely that one can make much progress
in computing the continuum path integral in proper-time gauge
explicitly, at least in four dimensions. Note that substantial
simplifications occur in the case $d\equ 3$, where there are no
further explicitly $\lambda$-dependent terms in the part of the
action indicated by the dots in formula (\ref{trunc}). In that case,
it remains to evaluate
\begin{equation}
Z_\epsilon^{d= 3}=\int [{\cal D}g_{ij}^\perp ]_\epsilon
\sqrt{ {\det}_{{\rm V}^{d-1}} ( 2\epsilon 
\bar\nabla_{0}^{\dagger}\bar\nabla_0) \;
{\det}_{\rm V}(\bar K^{\dag}\circ \bar K) }\  
\e^{-i\sqrt{-\epsilon} k^{ren} \bar S_{\epsilon} },
\label{zd3}
\end{equation}
with the action
\begin{equation}
\bar S_{\epsilon }^{d=3}=
\int dt \Bigl( 4\pi\chi +
\int d^2x \sqrt{\det \bar g_{ij}} \Bigl( -2\Lambda 
-\frac{\epsilon}{4} (\partial_{0}\bar g_{ij}) (\bar g^{ik}
\bar g^{jl}-\frac{1}{2} \bar g^{ij} \bar g^{kl})
 (\partial_{0}\bar g_{kl}) \Bigr)  \Bigr).
\end{equation}
Although this expression looks now tantalizingly simple, it is still
non-polynomial in the remaining metric components. We will not attempt 
here to evaluate (\ref{zd3}) further, but it would clearly be
interesting to relate it to any one of the known exact results
obtained in other approaches to three-dimensional quantum gravity
(see, for example, \cite{carlip}).

\section{Perturbative evaluation of the 3d path integral}

Although it was not our main motivation, one can take our 
formulation as the starting point for
a perturbative expansion around a given classical solution.
Depending on the solution one is interested in, the proper-time 
gauge is not necessarily the most convenient gauge choice
{\it perturbatively}. Also, we do not expect to find anything
new compared with previous calculations using other gauges.
Nevertheless, the calculation we will perform illustrates 
the general procedure outlined in the main part of the paper
and gives an explicit example of the cancellation mechanism at work.

We will study the path integral (\ref{jacact}) by computing its
perturbative expansion $Z^{(2)}$ around a particular classical solution.
For the sake of simplicity, we choose the spatial slices to
be flat two-tori (corresponding to a vanishing cosmological constant 
$\Lambda$), such that $M=[0,1]\times T^{2}$. There is a 
two-parameter set of classical solutions, 
\begin{equation}
g_{\mu\nu}=\left(\matrix{\epsilon & 0& 0\cr 
0 & V \frac{1}{\tau_{2}} &V \frac{\tau_{1}}{\tau_{2}}\cr
0 & V \frac{\tau_{1}}{\tau_{2}} & V
\frac{\tau_{1}^{2}+\tau_{2}^{2}}{\tau_{2}}\cr }\right),
\end{equation}
in a proper-time coordinate system $(t,x_{1},x_{2})$, 
where the $x_{i}$ are 
periodic and rescaled to have period 1 \cite{moncrief,carlip}. 
The metrics are parametrized by 
two modular parameters $\tau_{\alpha}$, and $V$ denotes a constant spatial 
area. We will perform our perturbative calculation around any one of the 
``straight torus solutions'' with $(\tau_{1},\tau_{2})=(0,\tau)$, 
$\tau > 0$, where 
the metric takes the diagonal form
\begin{equation}
g_{\mu\nu}^{0}=\mbox{diag }( \epsilon,\ V \frac{1}{\tau}, 
\ V \tau ).
\label{basemetric}
\end{equation}
Our starting point is the partition function
\begin{equation}
Z_\epsilon^{(2)}=\int [{\cal D}h_{ij}^\perp ]_\epsilon
\int [{\cal D}\delta\tilde\lambda ]_\epsilon\ J_\epsilon \tilde
J_\epsilon\ \e^{-\sqrt{\epsilon} S_\epsilon^{(2)} (h_{ij}^\perp,\delta
\tilde\lambda)},
\label{actpert}
\end{equation}
with the action given by
\begin{equation}
S_{\epsilon }^{(2)}=-\frac{\epsilon}{16\pi G}
\int dt \int d^2x \sqrt{\det g_{ij}}\ \left( 2\epsilon (\partial_0 \delta 
\tilde \lambda )^2 
-\frac{\epsilon}{4} (\partial_{0} h_{ij}^\perp )  g^{ik}
 g^{jl} (\partial_{0} h_{kl}^\perp ) \right) ,
\label{actpert12}
\end{equation}
and where both expressions are to be evaluated at the base metric 
(\ref{basemetric}). Since our end result will not depend on $\epsilon$
in a non-trivial way, we will from now on simply work with the
Euclidean value $\epsilon\equ 1$ and drop the subscripts $\epsilon$.
We are using the notation $h_{ij}=\delta g_{ij}$
for the tangent vectors at $g_{\mu\nu}^0\in \cal M$, 
and decompose them according to
(\ref{tangent}), (\ref{trace}). We will parametrize the directions
$h_{ij}^\perp$ perpendicular to the trace part explicitly as 
functions of the spatial vector fields $\xi_i$ and infinitesimal 
modular parameters $\delta\tau_\alpha$, such that
\begin{eqnarray}
h_{ij}^\perp &=& \nabla_i\xi_j +\nabla_j\xi_i -g_{ij} \nabla_k\xi^k 
+ \delta\tau_\alpha \langle \chi^{(\alpha)} ,\Psi_{(\beta)} \rangle
\ \delta^{\beta\gamma}\ \Psi_{(\gamma) ij}\nonumber\\
&=:& (\tilde L \xi)_{ij} 
+ \delta\tau_\alpha \langle \chi^{(\alpha)} ,\Psi_{(\beta)} \rangle
\ \delta^{\beta\gamma} \ \Psi_{(\gamma) ij},\;\;\;\;
\mbox{ with }
\chi^{(\alpha)}_{ij}:= \frac{\partial g_{ij}}{\partial\tau_\alpha}. 
\label{2ddecomp}
\end{eqnarray}

This variable change
is motivated by the standard decomposition of two-dimen\-sio\-nal
Riemannian metrics \cite{polya,carlip}
\begin{equation}
g_{ij}(\vec x,t)=\e^{2\lambda(\vec x,t)}f^*_{\vec x,t}\ 
\tilde g_{ij}(\vec x,t),
\label{2dmet}
\end{equation}
followed by a shift $\lambda \mapsto \tilde\lambda$ of the
conformal factor, c.f. (\ref{lamtilde}).
In the decomposition (\ref{2dmet}), $\tilde g_{ij}(\vec x,t)$ is one of 
a set of constant-curvature 
metrics, labelled by Teichm\"uller parameters $\tau_\alpha$, and $f$ a 
spatial diffeomorphism, with generators $\xi_i$. Note that as a
consequence of our gauge-fixing procedure, all quantities in
(\ref{2dmet}) carry an additional proper-time dependence.
To avoid misunderstandings,
let us also point out that the diffeomorphisms $f$
(which act in a standard way on the coordinates and 
$g_{ij}$'s, and map surfaces of constant $t$ into 
themselves) are of course no longer invariances of the gauge-fixed 
action.

The $\Psi_{(\alpha)ij}$ form a basis for $ker\ \tilde L^\dagger$, that is,
for the transverse traceless 
tensors, and we have chosen them to be orthonormal with respect to the
scalar product $\langle ,\rangle$ (which involves an integration over
the spatial directions only). Explicitly, they are given by
\begin{equation}
\Psi_{(1)}=\sqrt{\frac{V}{2}}
\left(\matrix{ 0&1\cr 1&0\cr }\right),\;\;\;
\Psi_{(2)}=\sqrt{\frac{V}{2}}
\left(\matrix{ \frac{1}{\tau}&0\cr 0&-\tau \cr }\right).
\end{equation}
Associated with the variable change $h_{ij}^\perp \mapsto
(\xi_i,\delta\tau_\alpha)$ is another Jacobian $\bar J$ 
which takes the form
\begin{equation}
\bar J = \sqrt{{\det}_{\rm V^2} (\tilde L^\dagger \tilde L)}\ 
\det \langle \chi^{(\alpha)} ,\Psi_{(\beta)} \rangle.
\end{equation}
A straightforward calculation yields 
\begin{equation}
{\det}_{\rm V^2} (\tilde L^\dagger \tilde L) = 
{\det}_{\rm V^2} (-2 g^{ij}\nabla_i\nabla_j )\equiv
{\det}_{\rm V^2} (-2 \Box )
\end{equation}
and
\begin{equation}
\det \langle \chi^{(\alpha)} ,\Psi_{(\beta)} \rangle =
-\frac{2 V}{\tau^2}.
\end{equation}
After the
coordinate transformation on the tangent space of metrics, 
the partition function reads 
\begin{equation}
Z^{(2)}=
J \tilde J \bar J \
\int [{\cal D}\xi_i ]
\int [{\cal D} \delta\tau_\alpha] 
\int [{\cal D}\delta\tilde\lambda ]\  \e^{- 
S^{(2)} (\xi,\delta
\tilde\lambda,\delta\tau)},
\label{actpert2}
\end{equation}
with the action 
\begin{equation}
S^{(2)}=
-\frac{V}{16\pi G}
\int dt \int d^2x \left( 2 (\partial_0 \delta 
\tilde \lambda )^2 +\frac{1}{2} g^{ij} \dot \xi_i \Box \dot\xi_j 
\right)+\frac{V}{16\pi G}\frac{1}{2\tau} 
\int dt\ (\partial_0\delta\vec\tau)^2 .
\end{equation}
We have pulled out the determinants from under the functional
integrations in (\ref{actpert2}), because they depend only on
the fixed background metric $g^0$. 

We can now perform the integrations over $\xi_i$, $\delta
\tilde\lambda$ and $\delta\tau$, since the action is quadratic 
in these variables.
Up to irrelevant (positive) constant terms,
the two integrations yield
\begin{eqnarray}
&&\mbox{ $\xi_i$-integral:  } \left({\det}_{\rm V^2}' (-\partial_0^2)
(-\Box ) \right)^{-\frac{1}{2}} \\
&&\mbox{ $\delta\tilde\lambda$-integral:  } \left({\det}_{\rm S}' 
(\partial_0^2) \right)^{-\frac{1}{2}}\\
&&\mbox{ $\delta\tau_\alpha$-integral:  } \left({\det}' 
(-\partial_0^2) \right)^{-1},\label{tauint}
\end{eqnarray}
where by definition all zero-modes have been excluded, and the
last determinant is that of a free two-dimensional particle
with mass $m=\frac{1}{16\pi G} \frac{V}{\tau}$.
The term coming from the $\delta\tilde\lambda$-integration
is of course the ill-defined determinant associated
with the conformal divergence.

A well-known subtlety arises in the evaluation of the determinant
of $(\tilde L^\dagger \tilde L)$ since the flat torus
possesses Killing vectors, leading to zero-modes of this
operator \cite{seriu,carlip,mottola}. This can be taken care
of by writing
\begin{equation}
{\det}_{\rm V^2} (\tilde L^\dagger \tilde L)=
V_K^{-1} {\det}_{\rm V^2}' (\tilde L^\dagger \tilde L),
\end{equation}
where $V_K$ denotes the (infinite) volume of the diffeomorphism
subgroup generated by the Killing vectors.
Since we do not keep track of {\it positive} constant and infinite factors,
we will simply drop this term. Also, it is not our aim here to 
investigate the possible physical significance of such zero-modes and 
we will remove them from now on wherever they occur.

Collecting all the determinants, the partition function is now
given by 
\begin{eqnarray}
Z^{(2)}&\simeq & J \tilde J \bar J'
\frac{1}{\sqrt{ {\det}_{\rm V^2}' (-\partial_0^2)\
{\det}_{\rm V^2}' (-\Box )\ {\det}_{\rm S}' (\partial_0^2)\  
({\det}'(-\partial_0^2))^2 }}
\nonumber\\
&=&\frac{\det \langle \chi,\Psi \rangle}{{\det}'(-\partial_0^2)}\ 
\frac{ \sqrt{ {\det}_{\rm S} 4(2+3C) \ 
{\det}_{\rm V}' (F\circ L)^\dagger (F\circ L) }}
{\sqrt{ {\det}_{\rm V^2}' (-\partial_0^2)\
{\det}_{\rm S}' (\partial_0^2) }}.
\label{penult}
\end{eqnarray}

The only non-trivial task remaining is the evaluation of the
Faddeev-Popov determinant ${\det}_{\rm V} (F\circ L)^\dagger (F\circ L)$.
At the metric $g^0 \in \cal M$, we can compute this determinant
explicitly, so there is no need to split off the time derivatives
as we did in (\ref{morefldet}). As a warm-up, let us
calculate the functional determinant $\det_{\rm 
V}(\nabla_{0}^{\dagger}\nabla_{0})\equiv \det_{\rm V} (-\partial_0^2)$.
In order not to have to deal with non-trivial boundary
terms we demand that the eigenfunctions on
$M=[0,1]\times T^2$ be periodic in the $x_1$- and 
$x_2$-direction, as well as in the time direction.
A complete set of eigenfunctions is then given by
\begin{equation}
{\vec\epsilon}^{\ (\kappa_1,\kappa_2,\omega,\mu)}= \vec\alpha^{(\mu )}\ 
\e^{i (\kappa_1 x_1+\kappa_2 x_2)}\ \e^{i\omega t},
\end{equation}
with
\begin{equation}
\vec\alpha^{(0)}=\sqrt{\frac{2}{V}}
\left( \matrix{1\cr 0\cr 0\cr }\right),\;
\vec\alpha^{(1)}= \sqrt{\frac{2}{\tau}}
\left( \matrix{0\cr 1\cr 0\cr }\right),\;
\vec\alpha^{(2)}=\sqrt{2\tau} \left( \matrix{0\cr 0\cr 1\cr }\right).
\end{equation}
The frequency $\omega$ takes discrete values 
$\omega=2\pi k$, $k=\pm 1,\pm 2,\pm 3,\dots$, and similarly the
$\kappa_i$ are given by
$\kappa_i=2\pi k_i$, $k_i=0,\pm 1,\pm 2,\dots$.
The eigenvectors are orthonormal with respect to the scalar product
(\ref{vecprod}), with
discrete eigenvalues 
\begin{equation}
\nu^{(\kappa_1,\kappa_2,\omega,\mu )}=\omega^{2}.
\end{equation}
The functional determinant is therefore the infinite product
\begin{equation}
{\det}_{\rm V} (-\partial_0^2)= 
\prod_{\omega,\kappa_1,\kappa_2} \omega^6.
\label{nabladet}
\end{equation}

Next we determine the eigenfunctions $\vec\eta$ of the Faddeev-Popov 
operator in 
\begin{equation}
[(F\circ L)^{\dag}(F\circ L)]_{\mu}{}^{\nu} \eta_{\nu} =\rho \eta_{\mu}.
\end{equation}
Making an ansatz of the form
\begin{equation}
\vec \eta = \vec\sigma (t) \e^{i \vec \kappa\cdot \vec x},
\end{equation}
for the three-vectors, 
one obtains a coupled set of eigenvalue equations, namely,
\begin{eqnarray}
\left( -4 \partial_0^2 +\frac{1}{V} (\tau \kappa_1^2 +
\frac{1}{\tau} \kappa_2^2 )\right) \sigma_0 -i \kappa_1 \frac{\tau}{V}
\partial_0\sigma_1 - i \kappa_2 \frac{1}{V\tau} \partial_0\sigma_2 
&=&\rho \sigma_0\\
 - i \kappa_1 \partial_0\sigma_0 -\partial_0^2\sigma_1 &=&\rho\sigma_1
 \label{kappa1}\\
 - i \kappa_2 \partial_0\sigma_0 -\partial_0^2\sigma_2 &=&\rho\sigma_2.
 \label{kappa2}
\end{eqnarray}
(We note in passing that it is clear from (\ref{kappa1},\ref{kappa2})
that in the presence of boundaries at $t=0,1$ it would not be
consistent to require a simultaneous vanishing of {\it all} components
of $\vec\sigma$ at the boundaries.)
One third of the eigenfunctions is easily found by setting
$\sigma_0=0$. The eigenvectors are of the form
\begin{equation}
\vec\eta^{\ (\kappa_1,\kappa_2,\omega,0)}= \left( \matrix{0\cr \frac{1}{\tau}
k_2\cr -\tau k_1\cr}\right) \e^{i \vec \kappa\cdot \vec x}\
e^{i\omega t},
\end{equation}
with eigenvalues
\begin{equation}
\rho^{(\kappa_1,\kappa_2,\omega,0)}= \omega^2.
\end{equation}
For the remaining eigenfunctions one finds  
\begin{equation}
\vec\eta^{\ (\kappa_1,\kappa_2,\omega,\pm)}=
\left( \matrix{ 1\cr
\frac{\omega}{\rho^\pm -\omega^2}\kappa_1  \cr
\frac{\omega}{\rho^\pm -\omega^2}\kappa_2 \cr }\right)
\e^{i \vec \kappa\cdot \vec x}\ e^{i\omega t},
\end{equation}
with the associated eigenvalues
\begin{equation}
\rho^{(k_1,k_2,\omega,\pm)}=
\frac{1}{2} \Bigl( 5\omega^2 +\frac{1}{V}(\tau \kappa_1^2+
\frac{1}{\tau}\kappa_2^2)\Bigr) \pm \frac{1}{2} 
\sqrt{\Bigl(5\omega^2 +\frac{1}{V}(\tau \kappa_1^2+
\frac{1}{\tau}\kappa_2^2)\Bigr)^2 -16 \omega^4}.
\end{equation}

For fixed $\kappa_1$, $\kappa_2$ and $\omega$, there are therefore 
three orthogonal eigenfunctions. 
The entire functional determinant
is thus given by
\begin{equation}
{\det}_{\rm V}' (F\circ L)^{\dag}(F\circ L)= 
\prod_{\omega\not= 0,\kappa_1,\kappa_2}\rho^{(\kappa_1,\kappa_2,\omega,0)}
\rho^{(\kappa_1,\kappa_2,\omega,+)}\rho^{(\kappa_1,\kappa_2,\omega,-)} 
\equiv
\prod_{\omega\not= 0,\kappa_1,\kappa_2} 4\omega^6.
\label{fpdet}
\end{equation}
This coincides (up to a constant factor) with the determinant of
the kinetic term we calculated earlier, that is,
${\det}_{\rm V}\nabla_0^\dagger\nabla_0$. Substituting the results for 
the determinants back into (\ref{penult}), we observe again 
an exact cancellation of the infinite products
{\it provided} that $C< -\frac{2}{3}$, in agreement with our
earlier non-perturbative results. 
Since the regularized determinant (\ref{tauint}) gives a term proportional
to $V/\tau$, our final result for the perturbative
partition function around flat torus metrics of the type
(\ref{basemetric}) is given by
\begin{equation}
Z^{(2)}=\frac{ \det \langle \chi,\Psi\rangle}
{ {\det}'(-\partial_0^2)}
\sim\frac{V^2}{\tau^3}.
\end{equation}

\section{Conclusions}

Inspired by recent attempts of constructing a non-perturbative 
propagator for gravity by discrete methods, we have investigated
the continuum gravitational propagator in proper-time gauge, concentrating
on the role played by the conformal mode of the metric.
Our starting point was the space of physical space-time metrics of
Lorentzian signature. After performing a generalized Wick rotation,
the partition function becomes real, but the Euclideanized action is 
seen to suffer from the usual ``conformal sickness'':
as a result of conformal excitations it is unbounded below. 

We then proceeded to determine the Faddeev-Popov determinants that
arise during the coordinate changes on the space of metrics, the first
from splitting off the gauge degrees of freedom associated
with the diffeomorphisms of the base manifold, the second from isolating
the part of the metric that has a negative-definite kinetic term.
Although an explicit evaluation of the functional determinants and
the non-perturbative path integral seem technically out of reach,
we showed that under certain assumptions about the behaviour of the
partition function under renormalization the conformal divergence 
in the action is cancelled by a corresponding Faddeev-Popov term in 
the measure. This conclusion also required that the signature of the
DeWitt metric was chosen to be indefinite, i.e. the constant $C$ in
the DeWitt measure (used to define inner products on the tangent space 
of metrics) satisfied $C< -\frac{2}{d}$ which we argued was a
natural condition. Our work can therefore be
seen as a non-perturbative generalization of earlier findings by
Mazur and Mottola \cite{mm} which -- although acknowledged by the authorities
\cite{pc} -- are maybe not widely appreciated.

Our results reinforce the evidence coming from dynamically triangulated 
formulations of quantum gravity \cite{ajl3d} that in a fully
non-perturbative path integral the conformal
mode is not a propagating degree of freedom and therefore the
conformal divergence is simply absent, contrary to what one may 
have expected from just looking
at the action or from considering reduced, cosmological models.
Although geometric configurations with large and negative 
action exist, they are effectively suppressed by the non-trivial
path-integral measure. This is an attractive proposition because 
it implies that the unboundedness of the action is no obstacle
in principle to the construction of a gravitational path integral.
Also, at a kinematical level (that is, before quantum gravity proper
has been solved), it
brings the covariant formulation of gravity into line with canonical 
treatments where the conformal degree of freedom is fixed by
imposing the Hamiltonian constraint. Although in the path integral 
this constraint is not enforced explicitly, it seems that 
the measure nevertheless {\it does} know about it. 

All our calculations were done in proper-time gauge, 
mimicking a similar procedure in the discrete Lorentzian gravity 
models of \cite{al,ajl1}. 
Note that there is nothing wrong in principle with choosing
a gauge in quantum gravity. Our choice of proper time does have
an invariant geometric meaning, but this by no means entails
that proper-time correlators assume a simple form in other gauges
or that any interesting physical quantity will be easily expressible 
in proper-time gauge.
This is simply an inevitable feature of gauge theories.
Obviously we expect our result about the absence of the
conformal sickness to be gauge-independent. In practice, an explicit
check may not be straightforward, since our construction of a
Wick rotation was closely tied to the proper-time gauge. 
There may be other gauge choices and other prescriptions of
Wick-rotating, but we are currently not aware of any concrete proposals.
As usual in quantum gravity, one is not exactly faced with an
embarrassment of riches when trying to quantize the theory. 

Clarifying the role of the conformal factor is only one step
in analyzing the properties of non-perturbative path integrals
for Lorentzian gravity. We do not expect that much progress can
be made in a continuum formulation in evaluating these 
quantities explicitly and showing 
how the non-renormalizability of the perturbative
approach is resolved non-perturbatively.
For a solution of these more fundamental problems we must
rely on the properly regulated discrete quantum gravity models that
are currently under investigation.

\vspace{.8cm}
\noindent {\it Acknowledgement.} We are grateful to J.\ Ambj\o rn
for discussion and comments. RL acknowledges the support of EU network
grant HPRN-CT-1999-00161.

\vspace{1cm}
\section*{Appendix 1}\label{gaugeapp}

In this appendix we show that every metric $g_{\mu\nu}$ in an 
infinitesimal neighbourhood of the constraint surface $\cal C$ can
be uniquely decomposed into an element in 
$\cal C$ and an infinitesimal diffeomorphism, parametrized by
a vector field $\xi_{\mu}$. Writing the original metric as
$g_{\mu\nu}=g_{\mu\nu}^{\rm pt}+h_{\mu\nu}$, we would like
to understand under which conditions the 
tangent vector $h_{\mu\nu}$ can be uniquely decomposed as
\begin{equation}
h_{\mu\nu} = h_{\mu\nu}^{\rm pt}+\nabla_{\mu}\xi_{\nu}
+\nabla_{\nu}\xi_{\mu},
\end{equation}
where $h_{\mu\nu}^{\rm pt}$ is tangential to $\cal C$, i.e.
of the form
\begin{equation}
h^{\rm pt}_{\mu\nu}=
\left( \matrix{0&\vec 0\cr \vec 0 & h_{ij}\cr } \right).\;\;
\label{coblock}
\end{equation}
This is tantamount to solving the set of equations
\begin{eqnarray}
2\nabla_{0}\xi_{0} &= &  h_{00} \label{dif1}\\
\nabla_{0}\xi_i + \nabla_i\xi_0 &= &  h_{0i} \label{dif2}
\end{eqnarray}
for $\vec\xi$, where the covariant derivatives refer to the
base point metric $g_{\mu\nu}^{\rm pt}$.
Since this metric is in proper-time gauge, its
Christoffel symbols take the form
\begin{eqnarray}
\Gamma_{00}^{\mu}&=&\Gamma_{0\mu}^{0}=0\nonumber \\
\Gamma^{0}_{lj}&=&-\frac12\ g_{lj,0} \nonumber \\
\Gamma^{i}_{0j}&= &\frac12\ g^{ik}g_{jk,0} \nonumber\\
\Gamma^{i}_{lj}&= &\frac12\ g^{ik}\left(g_{lk,j} + g_{jk,l} - g_{lj,k}\right).
\end{eqnarray}
Using these explicit expressions we obtain
\begin{eqnarray}
\xi_{0}(t,{\bf x}) &=& \frac12 \int_0^t  dt'\ h_{00}(t',{\bf x}) \\
\partial_t \xi_i + a_{ij}\xi_j & = & b_i(t,{\bf x}),
\end{eqnarray}
where $b_i\equ h_{0i}- \partial_i\xi_0$, and
$a_{ij}\equ - 2\Gamma^i_{0 j}$.
The system of differential equations determining $\xi_i$ has unique
solutions if 
\begin{itemize}
\item[1)] $b_i$ and $a_{ij}$ are smooth and continuous in the interval of 
interest, namely, [0,t];
\item[2)] the $a_{ij}$ satisfy a Lipschitz condition 
\begin{equation}
\sum_{i=1}^{d-1} |a_{ij}|< k_j,
\label{bounded}
\end{equation}
where the $k_j$ are arbitrary constants, greater than zero.
\end{itemize} 
These conditions are satisfied for all metrics
$g_{\mu\nu}^{\rm pt}$, which we assume to be non-degenerate and
sufficiently differentiable. The boundedness property (\ref{bounded})
follows from the compactness of the base space
${}^{(d)}M=[0,t]\times {}^{(d\mi 1)}\Sigma$.

\end{document}